\documentclass[12pt,preprint]{aastex}
\begin{document}
\def\msun{M$_\odot$}

\title{Microlensing towards the Magellanic Clouds: Nature of the Lenses
and Implications for Dark Matter}

\author{Kailash C. Sahu} \affil{Space Telescope Science Institute,
Baltimore, MD 21218} 

\begin{abstract} 
A close scrutiny of the microlensing results towards the Magellanic
clouds reveals that the stars within the Magellanic clouds are major
contributors as lenses, and the contribution of MACHOs to dark matter 
is 0 to 5\%. The principal results which lead to this conclusion are
the following. 

\noindent (i) Out of the $\sim$17 events detected so far towards the  Magellanic 
Clouds, the lens location has been securely determined  for one 
binary-lens event through its caustic-crossing timescale. In this case,
the lens was found to be  within the Magellanic Clouds. Although less
certain, lens locations have been  determined for 3 other events  and
in each of these 3 events, the lens is most likely within the
Magellanic clouds. 

\noindent (ii) If most of the lenses are MACHOs in the Galactic  halo, the
timescales would imply that the MACHOs in the direction of the LMC have
masses of the order of 0.5 \msun, and the MACHOs in the
direction of the SMC have masses of the order of 2 to 3 \msun.
This is inconsistent  with even the most flattened model of the Galaxy.
If, on the other hand,  they are caused by stars within the Magellanic
Clouds, the masses of the stars are of the order of 0.2 \msun \ 
for both the LMC as well as the SMC. 

\noindent (iii) If 50\% of the lenses are in binary systems similar to
the stars in the solar neighborhood, $\sim$10\% of the events are
expected to show binary characteristics. The fact that the two observed
binary events are caused by lenses within the Magellanic Clouds would
imply that there should be a total of $\sim$20 events caused by lenses
within the Magellanic Clouds. This implies that most of the
microlensing events  observed so far are probably caused by stars
within the Magellanic Clouds.

\noindent (iv) If  the microlensing events are caused by MACHOs of 0.5
\msun, as claimed from the LMC observations, about 15 events should
have been detected by now towards the SMC, with timescales of $\sim$40
days. The fact that no event has been detected towards the SMC caused
by MACHOs (both the events detected towards the SMC have been shown to
be due to self-lensing) places severe constraints on the MACHO
contribution and suggests that the contribution of MACHOs to dark
matter is consistent with zero, with an upper limit of 5\%.

\end{abstract} 

\section{Introduction}

Most spiral galaxies are known to have flat rotation curves (e.g.
Begeman et al. 1991). In the outer parts of the galaxies in particular,
the visible matter falls far short of what is required to explain the
flat rotation curves. In our own Galaxy, the rotation curve is observed
to be flat up to at least 16 kpc from the center (e.g. Fich et al.
1989). The scenarios proposed to explain the flat rotation curves are
either departure from Newtonian dynamics in large scale, or presence of
dark matter (for a review see  Sanders 1990). In the latter and more
conventional scenario, the rotation curves would imply that a
significant part of the mass of the galaxies, including our own,
resides in the halo in some form of dark matter. The nature of this
dark matter has been hypothesized to be in either of the two forms:
MACHOs (i.e. massive compact halo objects which is a collective term
for `Jupiters', brown dwarfs, red dwarfs, white dwarfs, neutron stars
or black holes); or elementary particles such as massive neutrinos,
axions or WIMPs (weakly interacting massive particles such as photinos
etc.). One important distinguishing factor of the MACHOs is that they
can have detectable gravitational effects and can cause ``gravitational
microlensing" of background stars.
 
Einstein (1936) was the first to point out that a star can act as a 
gravitational lens for another background star, if the two are
sufficiently  close to each other in the line of sight. Given the
observational  capabilities of that time, Einstein had however
considered this to be a purely theoretical exercise since there was
``no hope of observing such a phenomenon directly". 

Paczy\'nski (1986) worked out the probability of such  microlensing
events by MACHOs and showed that if the halo of our Galaxy is made up
of  MACHOs, the probability of finding them is 5 $\times$  10$^{-7}$,
independent  of their mass distribution. He suggested an experiment to
look for such  events using the LMC stars. Such an experiment was taken
up by 2 groups  who reported their first results in 1993 (Alcock et al.
1993;  Aubourg et al.  1993). In 6 years' monitoring of millions of
stars towards the Magellanic Clouds, so far about 17 events have been
detected towards the Large Magellanic  Cloud (LMC) and 2 events towards
the Small Magellanic Cloud (SMC). 

The observed microlensing optical depth as derived from these
observations is $\sim 1 \times 10^{-7}$. This is about a factor of 5
lower than what would be expected if the dark matter is entirely made
up of MACHOs. To complicate the issue further, the simple microlensing
light curves cannot tell us the location of the lenses because of a
degeneracy between the distance, mass and the proper motion of the
lens. So the lenses can be stars in the Galactic disk (Gould ~1994, 
Evans et al. 2001, Gates and Gyuk, 2001),  MACHOs in the Galactic halo 
(Alcock et al. 2000a), or stars within the LMC itself (Sahu 1994a,b;
Evans and Eamonn 2000).  As a result, the interpretation of these
lensing events, and their locations in particular, have remained
uncertain. 

\section{Overview of the Problem}

The contribution of known stars to the microlensing optical depth had
not been calculated until after the first microlensing events were
reported towards the LMC. Such calculations revealed  that the
microlensing optical depth due to the known stars within the LMC is
close to the observed optical depth, particularly in the region of the
bar where the microlensing event was detected (Sahu 1994a,b; Wu 1994).
This was used to argue that the stars within the Magellanic Clouds must
play a significant role as lenses. 

Lensing by stars within the Magellanic clouds is now commonly referred
to as ``self-lensing". It is known for a long time (e.g. Schneider et
al. 1992), and it was re-derived by Gould (1995) that the self-lensing
microlensing optical depth can be expressed as a function of the
velocity dispersion 

\begin{equation} \tau = 2 \times sec^2 (\theta) v^2/c^2 \end{equation}

\noindent where $\theta$ is the inclination of the disk of the LMC (the
best estimated value of which is $34.7\pm 6\deg$, van der Marel and
Cioni, 2001). Since the velocity dispersion of the stars in the disk of
the LMC is about 20 km/s, the above equation would imply that the
self-lensing optical depth is about $2 \times 10^{-8}$, which is too
low to explain the observed microlensing events.

However, there are some problems in such an approach. First, eq. (1)
can be applied only for a virialized system. The Magellanic Clouds are
not relaxed, and far from virialized systems: they are dynamically
disturbed, they show tidal-tail structures, and the velocity
dispersions are different for low and high-mass objects (for a review,
see Westerlund 1997). Second, the lenses are low-mass objects for
which the velocity dispersions are unknown. Third, the velocity
dispersions in the region of the bar, where most of the events have
been found so far, is also unknown.

A cautionary note seems appropriate here. A similar story had repeated
towards the Galactic bulge. The velocity dispersion of the Galactic
bulge stars would imply a microlensing optical depth which is a factor
of at least 3 smaller than the observed optical depth.  In the case of
the Galactic bulge, the re-discovery of the Galactic bar, with an
inclination of 15$^\circ$ to the line of sight helped in solving the
puzzle of the discrepancy between the observed and the expected
microlensing optical depth (Kiraga and Paczy\'nski, 1994; Paczy\'nski
et al. 1995).

In the meantime, several papers have appeared with detailed
calculations of the optical depth taking into account the extra effects
such as `diffusion' of small-mass objects, which causes the velocity
dispersion of less massive objects to be larger (e.g. Salati et al.
1999). 

The full range of the calculated optical depth due to the stars within
the LMC, as published in the literature so far, is $1 \times 10^{-8}$
to $ 2 \times 10^{-7}$ (e.g. Gyuk et al. 2000, Graff 2001, Salati et
al. 1999, Aubourg et al. 1999). The high end of this self-lensing
optical depth would imply that all the observed events are caused by
stars within the LMC and that the contribution of MACHOs to the dark
matter is negligible. The low-end of the self-lensing optical depth
would imply that none of the observed events are caused by stars within
the Magellanic clouds, and hence the events are caused most likely by
MACHOs in the halo. In the extreme case where all the events are caused
by MACHOs in the Galactic halo, the implied contribution of MACHOs to
dark matter is about 25\% (Alcock et al. 2000a, Afonso et al. 2002). 

Cleary, it is important to know the exact location and nature of the
lenses, which can shed some light on the whether MACHOs can account for
the long-sought dark matter in the halo. We obviously need some
observational tests to guide us in resolving this puzzle. The rest of
the paper mainly deals with such tests and the current status of the
results from these tests.

It is worth pointing out here that, in the self-lensing hypothesis,  it
is implicit that the sources and the lenses reside within the LMC.  The
sources cannot be considered as background sources since the derivation
of the observed microlensing optical depth assumes that all the
monitored stars  which are {\it within} the Magellanic clouds
contribute to the microlensing optical depth.

\section{Observational Tests}

The simple microlensing light curve is inadequate to the
determine the lens location because of a degeneracy between the
distance, the proper motion and the mass of the lens.  A typical
difficulty is that, in order to determine the lens location, both the
mass and the proper motion of the lens are needed.  Generally, these
quantities can only be estimated statistically, since the lens is not
detected directly.  As a result, the  interpretations of these lensing
events, and the lens locations in particular, remain uncertain. 

Fortunately, there are several observational tests which can be used to
infer the nature of the lenses, including some tests which can
be used for direct determinations of their locations. The
observational tests that I will discuss here, some of which have
already provided clear results, are the following:\\

{\parskip 0pt
\parindent 0pt
(i) Determination of lens location through ``caustic-crossing" timescale\\
(ii) Other direct determinations of lens locations\\
(iii) Frequency of observed binary-lens events\\
(iv) Timescales of SMC vs. LMC events\\
(v) Spatial distribution of the microlensing events}

\subsection{``Caustic-crossing" timescale}

Clearly, the best test would be to measure the location of the lens
directly. In a few special cases, such a direct determination of the
lens location is indeed possible. One such special case is when the
lens is a binary, and the source crosses the caustics produced by the
binary lens. Since a caustic is essentially a straight line in space,
the time taken by the caustic to cross the source provides a direct
measure of the proper motion of the lens projected onto the source
plane.  If the lens is in the halo at a distance of  $\sim$10 kpc, then
the expected proper motion of the lens is about 200 km/s which,
projected onto the source plane, is about 1000 km/s. So the time for
the caustic to cross the source would be of the order of half an hour.
If, on the other hand, the lens is within the Magellanic Clouds, the
expected proper motion is about 50 km/s, and hence the caustic crossing
time is expected to be of the order of 10 hours. Thus, monitoring a
caustic crossing provides a powerful method to determine the location
of the lens.  

\subsubsection{MACHO 98-SMC-1}

Such an opportunity presented itself after the first binary-lens event,
MACHO 98-SMC-1, was discovered by the MACHO  collaboration in 1998
(Alcock et al. 1999). After the first caustic crossing was reported,
the PLANET collaboration, with its 24-hour access to telescopes around
the world, predicted a second caustic crossing and began monitoring
this event, with a particular interest in fully sampling the second
caustic crossing (Albrow et al. 1999). The time for the caustic to
cross the source was found to be ~8.5 hours (Fig. 1), which
demonstrated that the lens is within the SMC (Fig. 2). The result was
further confirmed by the EROS, MACHO and MPS collaborations, who also
concluded that the lens is most likely within the SMC (Afonso et al.
1998, 2000, Alcock et al. 1999, Rhie et al. 1999). 

\begin{figure} \plotone{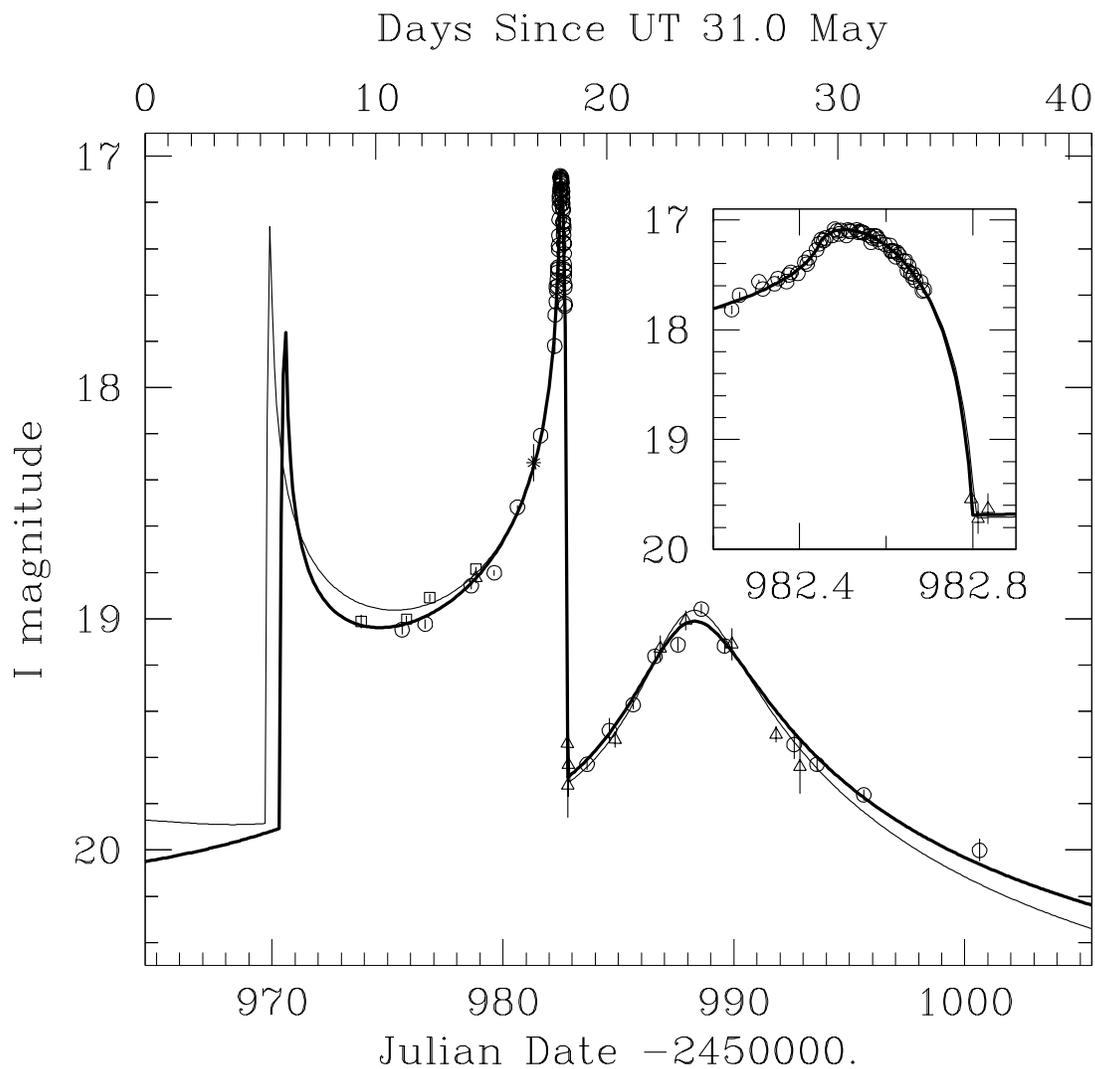} \caption{Light curve of the
PLANET data for MACHO-98-SMC-1.  Shown are the data from the SAAO 1 m
({\it circles}), the CTIO 0.9 m ({\it squares}), the  CTIO-Yale 1 m
({\it triangles}), and the Canopus 1 m ({\it asterisks}).   The inset
covers about 0.6 days, corresponding to less than one tick mark on the
main figure. (Taken from Albrow et al. 1999)} \end{figure}

\begin{figure} \plotone{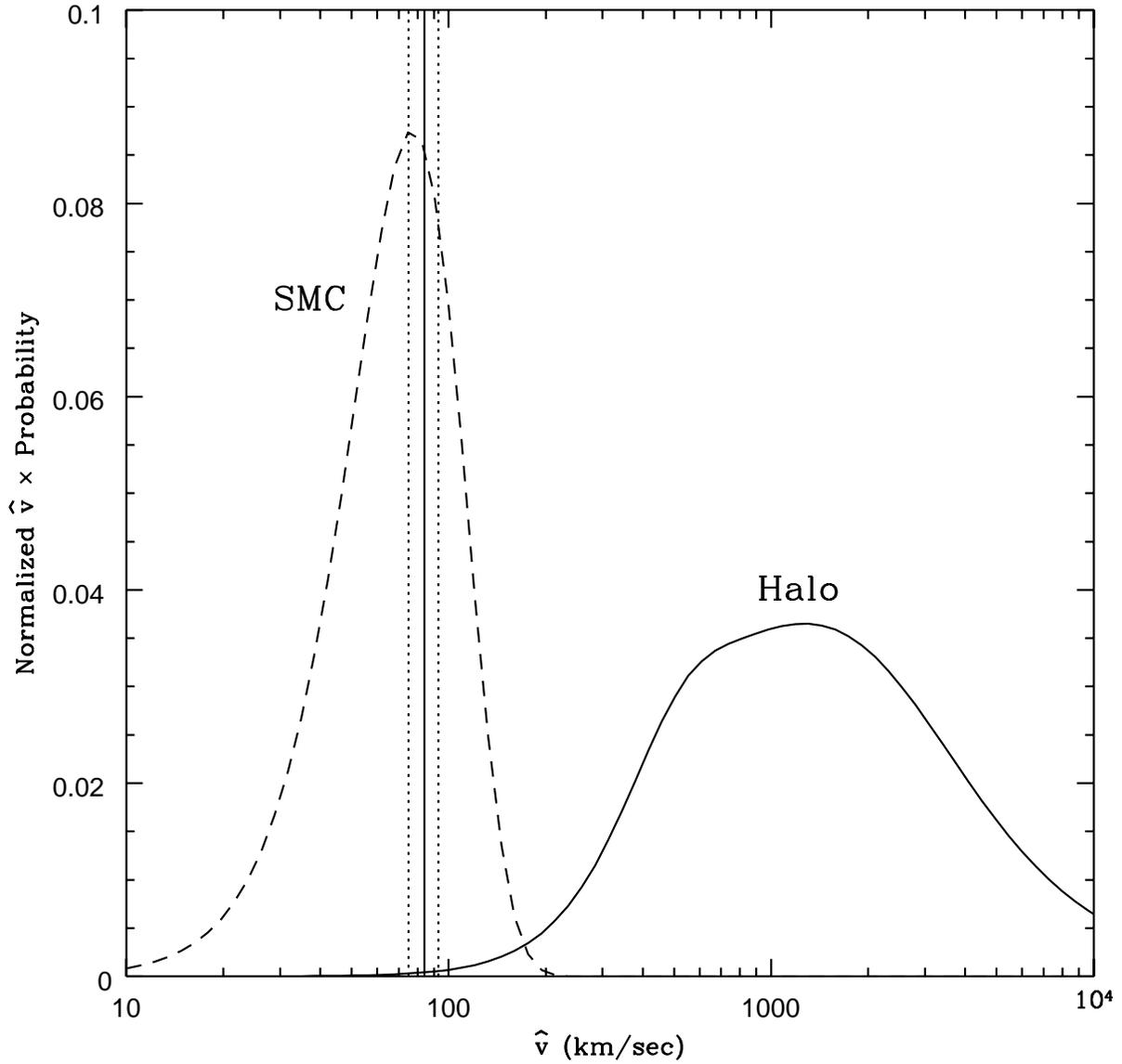} \caption{Predicted velocity
distributions (at the source plane) for halo and SMC lenses. The
measured value of the velocity and error bars are shown as vertical
lines. The observed value is inconsistent with the halo hypothesis. 
The lens and source could possess normal SMC kinematics. (Taken from
Alcock et al. 1999)} \end{figure}

\subsubsection{MACHO-LMC-9}

The second binary-lens event, MACHO-LMC-9, was discovered by the MACHO
collaboration towards the LMC, for which the observations are not as
dense (Alcock et al. 1997a). The caustic crossing timescale of the
event also suggests that the lens is most likely within the LMC (Fig.
3), although the sparse sampling of this  event quality has led to some
doubt as to whether this is a microlensing event at all (Alcock et al.
2000a,b). 

\begin{figure} \plotone{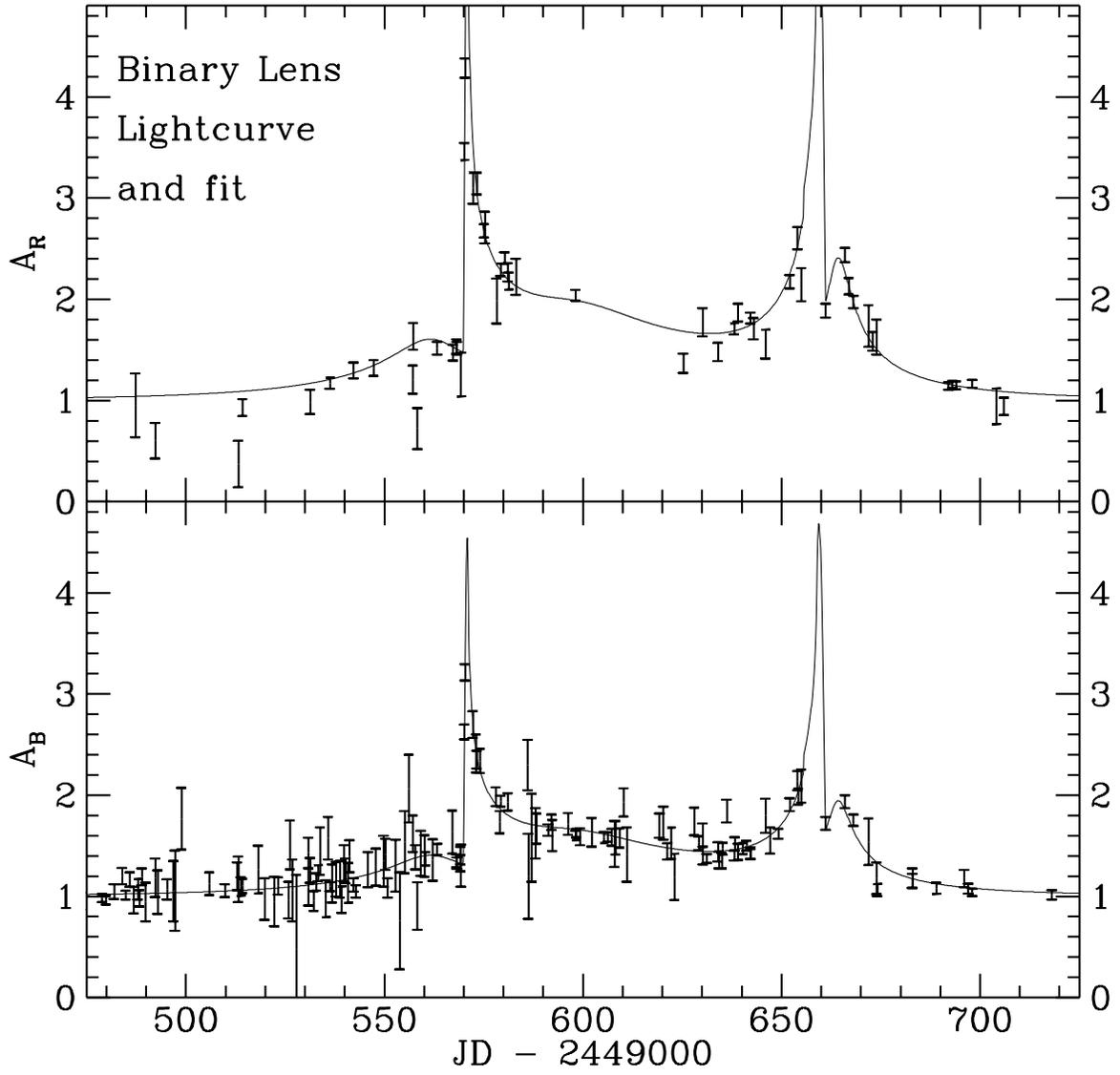} \caption{Light curve of LMC-9 in the
around the peak of the event, and the best-fit binary lens light curve.
The most likely lens location is within the LMC. (Taken from 
Alcock et al. 1997a).} \end{figure}

\subsection{Other determinations of lens locations}
\subsubsection{MACHO 97-SMC-1}

97-SMC-1 is the first microlensing event observed towards the SMC,
which has a large timescale of 220 days, in contrast to the average
timescale of 40 days as observed for the more than dozen microlensing
events towards the LMC (Alcock et al. 1997b).  

\begin{figure} \plotone{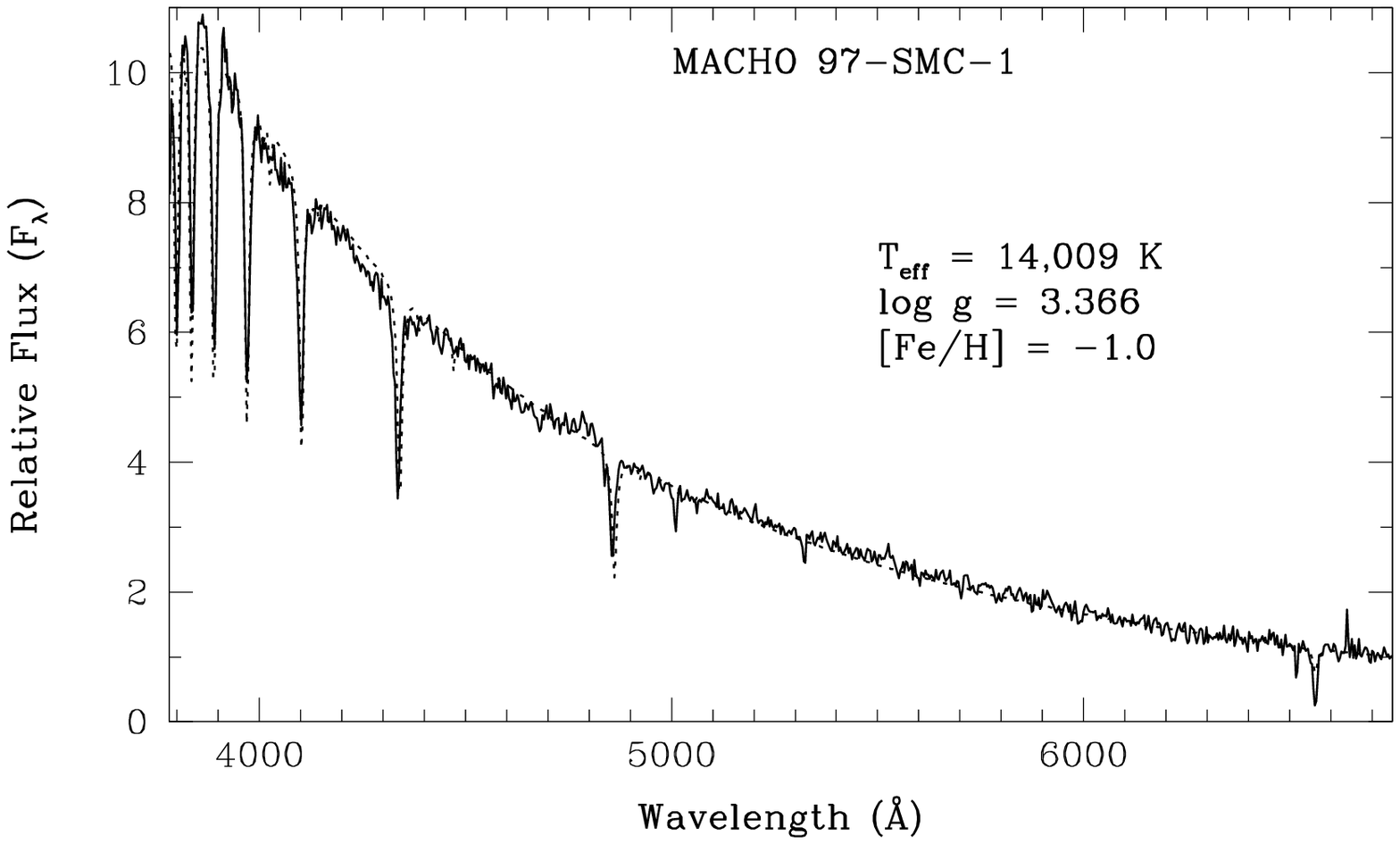} \caption{Observed spectrum of
MACHO 97-SMC-1 taken on  30th May 1997, just after the source had
crossed the Einstein ring radius of the lens (solid curve), along with
the best fit stellar model spectrum (dashed line). The contribution
from the lens is negligible which, combined with the microlensing time
scale,  implies that the lens must be within the SMC. (Taken from Sahu
and Sahu, 1998).} \end{figure}

If the lens is a MACHO in the Galactic halo, the timescale of the event
would suggest that the mass of the lens is about 3 \msun . Clearly,
this cannot be a normal star of 3 \msun \  since it would be too bright
and would be easily detectable. Indeed, a spectrum of the source
obtained by Sahu and Sahu (1999) shows the spectrum to be a pure B-type
spectrum  as expected from a source within the SMC, without any
contribution from the lens (Fig. 4). This would suggest that the lens
is either a black hole in the Galactic halo, or a small mass object
within the SMC. 

A black hole in the Galactic halo is an unsatisfactory explanation
since that would imply that the MACHOs in the direction of the LMC are
predominantly 0.5 solar mass objects and the MACHOs in the direction of
the SMC are predominantly 3 solar mass objects. The locations of the
LMC and the SMC too close to permit such a mass segregation in these
two directions even in the  most extreme flattened model of the Galaxy.
This leaves us with the only other alternative that the lens must be a
small mass object within the SMC. 

Furthermore, as explained in the next section, this long timescale of
the SMC events is precisely what one would expect if the lenses are
within the Magellanic Clouds, which further supports the view that the
lens is most likely within the SMC.

\subsubsection{MACHO 96-LMC-2} 

Lens locations have been estimated for another event MACHO 96-LMC-2,
where the source is a binary (Alcock et al. 2001a). In this case too,
the lens is most likely within the LMC.

Thus there are 4 cases so far, for which we know the locations of  the
lenses. In all the cases, without exception, the most likely locations
of the lenses are within the Magellanic Clouds.

It is worth noting here that the lens location has been recently
determined for an event caused by a star in the local Galactic disk
(Alcock et al. 2001b). The events caused by nearby lenses are expected
to exhibit ``parallax" effects which are easier to detect. The fact
that only one event has been detected caused by a star in the local
Galactic disk is consistent with the expected frequency of nearby stars
acting as lenses. 

\subsection{Frequency of observed binary-lens events}

For the sake of argument, let us start with the assumption that  all
the events except the two binary events for which we know that the
lenses are within the Magellanic clouds are caused by MACHOs in the
halo. This would be quite extra-ordinary since this would mean that the
only two events that are caused by the stars within the Magellanic
clouds not only happen to be binaries, but they also show caustic
crossing structure in their light curves. If one event caused by a star
in the Magellanic clouds is observed, the probability that the lens
would be a binary is about 50\% (assuming 50\% of the lenses are in
binary systems similar to the stars in the solar neighborhood). Since
only 20\% of the binary-lens events are expected to {\it show}
caustic-crossing structures (which is borne out by the observations
towards the Galactic bulge as well), the probability that it would {\it
show} a caustic-crossing structure is 10\%. The probability that a
second event caused by a Magellanic cloud star would again be a binary
and show caustic structure is again 10\%. The probability that two
successive events caused by Magellanic cloud stars will both be
binaries and show caustic crossing structures is thus 1\%, and hence
extremely unlikely. We can follow the same argument in reverse and ask
the following question: if 2 events which show binary characteristics
are within the Magellanic clouds, how many of the observed events
should be caused by stars within the Magellanic clouds? Since 10\% of
the events are expected to show binary characteristics, one would
expect about 20 events to be caused by stars within the Magellanic
clouds. This is more or less the number of events observed, which
suggests that most of the lenses must be within the Magellanic clouds.

\subsection{Timescales of SMC vs. LMC events}

The relative timescales of the LMC and the SMC events provide further
indications on the locations of the lenses. If the lenses are MACHOs in
the Galactic halo, the characteristics of the lenses towards the LMC
and the SMC are expected to be the same and hence both the LMC and the
SMC events are expected to have similar timescales. On the other hand,
if the lenses are within the Magellanic Clouds, the expected timescales
of the LMC and the SMC events are expected to be very different. The
LMC has a depth of less than 1 kpc in a typical line of sight whereas
the SMC as a  much larger depth of about 5 kpc along a typical line of
sight. So the typical distance between the lens and the source is much 
smaller in the case of LMC than in the case of the SMC. As a result,
the size of the Einstein ring (which scales as the square root of the
distance between the source and the lens) is much smaller for the LMC
events than than for the SMC events. Since the velocity dispersions of
the stars within the LMC and the SMC are similar, the timescales of the
SMC events are expected to be much longer than the LMC events, if the
lenses are within the Magellanic Clouds.  

\begin{table*} {\caption{Lens masses for different scenarios}}
\begin{tabular}{ccccc} \\ \hline

Line of & \quad No. of \quad & \quad M$_{lens}$ (halo) \quad & \quad M$_{lens}$ 
(LMC/SMC) \quad & \quad M$_{lens}$ (local disk) \quad    \\
Sight & \quad Events \quad &($M_\odot$) & ($M_\odot$) & ($M_\odot$) \\
\hline

LMC & $\sim$ 17 & $\sim$0.5 & $\sim$0.2 & $\sim$0.5\\
SMC & $\sim$ 2 & $> 2$ & $\sim$0.2 & $\sim$2 \\

\hline
\end{tabular}
\end{table*}
\clearpage

Two events have been observed towards the SMC so far which have 
timescales of 75 and 125 days. This would correspond to lens masses of
2 to 3 solar masses if the lenses are in the halo, and about 0.2 to 0.3
solar masses if the lenses are within the Magellanic clouds. For the
$\sim$17 events observed towards the LMC, the timescales are much
shorter, which would correspond to lens masses of $\sim$0.5 solar
masses if the lenses are in the halo, and $\sim$0.2 solar masses if the
lenses are within the LMC. This is shown it tabular form in Table 1. If
the events are predominantly caused by MACHOs in the halo, one faces
the unrealistic consequence that the MACHOs in the direction of the LMC
and the SMC are of very different mass. Self-lensing is the only
scenario which gives consistent masses for both the LMC and the SMC
events. The longer durations of the SMC  events are a natural
consequence of the self-lensing hypothesis.

\subsection{Spatial distribution of the events}

Finally, the spatial distribution of the events provide some extra
insight into the nature of the lenses. As first pointed out by Sahu
(1994 a,b), the events should be more concentrated towards the bar of
the LMC if the lensing is caused by the stars within the LMC. If the
events are caused by MACHOs then  the events (for a given number of
monitored stars) should be uniformly distributed over the whole of the
LMC. Unfortunately, all the analyses by the MACHO group so far have
been confined to the region around the bar of the LMC. The EROS group,
however, have mostly monitored in the region outside the bar, and have
not detected any event. From their non-detection of microlensing
events, they derive a microlensing optical depth which is much smaller
than the optical depth determined by the MACHO group. This is
consistent with the fact that stars within the Magellanic clouds play a
dominant role as lenses. 

Monitoring some regions far from the bar of the LMC to look for
microlensing events would provide a clear test on whether the lensing
events are caused by MACHOs or stars within the LMC (e.g. Stubbs,
1999).

\section{Implications on Dark Matter}

The most significant insight into the contribution of dark matter, in
my view, comes from a comparative study of the number of observed
events towards the LMC and the SMC, and their timescales. So far, 2
microlensing events have been detected towards the SMC with time scales
of 75 and 125 days.  From these two events, the optical depth towards
the SMC has been estimated to be $\sim$2 $\times$ $10^{-7}$ which is
about the same as the optical depth towards the LMC (Alcock et al.
1999). In terms of optical depth, these two events are equivalent to
about 15 events with timescales of $\sim$40 days (optical depth scales
as the square of the timescale). Thus,  if the microlensing events are
caused by 0.5 solar mass MACHOs as claimed from the LMC observations,
about 15 such events should have been detected towards the SMC by now. 
Furthermore, as discussed earlier, both of the observed SMC events have
been shown to be caused by stars within the SMC. Thus no microlensing
event caused by a MACHO has been detected towards the SMC, whereas 15
such events should have been detected if the contribution of MACHOs to
dark matter is indeed 20\% as claimed from the detection of $\sim$20
microlensing events towards the LMC. This leads us to the inevitable
conclusion that the contribution of MACHOs to dark matter is less than
2\%, with a strong upper limit of 5\%. 

\section{Conclusions}

Results from several observational tests are already available which
can be used to clearly distinguish between different scenarios for the
nature of the lenses. I have argued that the results obtained so far
point to the fact that the observed microlensing events towards the
Magellanic clouds are caused by the stars within the Magellanic Clouds.
Consequently, the contribution of MACHOs to the dark matter is
consistent with zero, with a strong upper limit of 5\%. However, some
estimates suggest that the known stars within the Magellanic Clouds
fall short by a factor of a few to explain the observed microlensing
events. The contribution to the microlensing optical depth by the known
stars is, however, very uncertain, and further work would be needed to
resolve this issue.   

\bigskip

{\parindent=0pt \parskip=0pt {\bf{References:}}
\small skip

Afonso, C., et al. 1998, A\& A, 337, L17

Afonso, C., et al. 2000, ApJ, 532, 340

Afonso, C., et al. (EROS Collaboration) 2002, A\& A, in press
(astro-ph/0212176)

Albrow, M. et al. 1999, ApJ, 512, 672

Alcock, C. et al. 1993, Nature, 365, 621

Alcock, C. et al. 1997a, ApJ, 491, L11

Alcock, C. et al. 1997b, ApJ, 486, 697

Alcock, C. et al. 1999, ApJ, 518, 44

Alcock, C. et al. 2000a, ApJ, 542, 281 

Alcock, C. et al. 2000b, ApJ, 541, 270

Alcock, C. et al. 2001a, ApJ, 552, 259

Alcock, C. et al. 2001b, Nature, 414, 617

Aubourg, E. et al. 1993, Nature,  365, 623

Aubourg, E. et al. 1999, A\& A, 347, 850

Begeman, K.G., Broeils, A.H., Sanders, R.H., 1991, MNRAS, 249, 523

Einstein, A., 1936, Science, { 84}, 506 

Evans, N. W., Gyuk, G., Turner, M. S., Binney, J., 1998, ApJ, 501, L45

Evans, N. W., Eamonn, K., 2000, ApJ, 529, 917 

Fich, M, Blitz, L, Stark, A., 1989, ApJ, { 342}, 272

Gates, E.L., Gyuk, G., 2001, ApJ, 547, 786

Gould, A. 1995, ApJ, 441, 77

Gould, A., 1994, ApJLett, 421, L71

Graff, D., 2001, Microlensing 2000: A New Era of Microlensing
Astrophysics, ASP Conference Proceedings, Vol. 239. Edited by J. W.
Menzies and Penny D. Sackett, p.73

Gyuk, G., Dalal, N., Griest, K.,  2000, ApJ, 535, 90

Kiraga, M., Paczynski, B., 1994, ApJ, 430, L101

van der Marel, R., Cioni, M.L., 2001, AJ, 122, 1807

Paczy\'nski, B, 1986, ApJ, { 304}, 1

Paczynski, B., Stanek, K. Z., Udalski, A., Szymanski, M., Kaluzny, J.,
Kubiak, M., Mateo, M., Krzeminski, W.,  1994, ApJ, 435, L113 

Palanque-Delabrouille, N., et al. (EROS collab) 1997, Astron.
Astrophys. 332, 1

Palanque-Delabrouille, N.,  2001, New Astron. Rev., 45, 395

Rhie, S.H. et al. 1999, apJ, 522. 1037

Sahu, K.C. 1994a, Nature, 370, 275 

Sahu, K.C. 1994b, PASP, 106, 942

Sahu K.C., Sahu, M.S., 1998, ApJ, 508, 147

Salati, P. et al. 1999, A\& A, 350, L57

Sanders, R.H., 1990, Astron. Astrophys. Rev., { 2}, 1

Schneider, P, Ehlers, J. and Falco, E.E., 1992, ``Gravitational
Lensing", published by Springer-Verlag.

Stubbs, C, In The Third Stromlo Symposium: The Galactic Halo, eds.
Gibson, B.K., Axelrod, T.S. \& Putman, M.E., ASP Conference Series Vol.
165, p503

Westerlund, B.E., 1997, ``The Magellanic Clouds", Cambridge Univ. Press

Wu, X-P. 1994, ApJ., { 435}, 66

\end{document}